\documentclass[graybox]{svmult}

\usepackage{mathptmx}       % selects Times Roman as basic font
\usepackage{helvet}         % selects Helvetica as sans-serif font
\usepackage{courier}        % selects Courier as typewriter font
\usepackage{type1cm}        % activate if the above 3 fonts are
\usepackage{makeidx}         % allows index generation
\usepackage{graphicx}        % standard LaTeX graphics tool
\usepackage{multicol}        % used for the two-column index
\usepackage[bottom]{footmisc}% places footnotes at page bottom

\makeindex             % used for the subject index
\begin{document}

\title*{The Acceleration Scale, Modified Newtonian Dynamics, and Sterile Neutrinos}
\author{Antonaldo Diaferio and Garry W. Angus}
\institute{Invited review contribution to {\it Gravity: Where Do We Stand?} edited by R. Peron, V. Gorini and U. Moschella, \copyright\ Canopus Academic Publishing Limited, in press.\\
\\
Antonaldo Diaferio\at Dipartimento di Fisica, Universit\`a degli 
Studi di Torino, and Istituto Nazionale di Fisica Nucleare (INFN), Sezione di Torino, 
Via P. Giuria 1, 10125, Torino, Italy, \email{diaferio@ph.unito.it}
\and Garry W. Angus\at Astrophysics, Cosmology and Gravity Centre, University of Cape Town, Private Bag X3, Rondebosch, 7700, South Africa, \email{angus.gz@gmail.com}}
\maketitle

\abstract{General Relativity is able to describe the 
dynamics of galaxies and larger cosmic structures only if
most of the matter in the Universe is dark, namely it 
does not emit any electromagnetic radiation. Intriguingly, 
on the scale of galaxies, there is strong observational 
evidence that the presence of dark matter appears to be necessary 
only when the gravitational field inferred from 
the distribution of the luminous matter falls 
below an acceleration of the order of $ 10^{-10}$~m~s$^{-2}$. 
In the standard model, which combines Newtonian gravity with 
dark matter, the origin of this acceleration scale 
is challenging and remains unsolved. On the contrary, the full set of
observations can be neatly described, and were partly predicted, by
a modification of Newtonian dynamics, dubbed MOND, 
that does not resort to the existence of dark matter.
On the scale of galaxy clusters and beyond, 
however, MOND is not as successful as on the scale of galaxies, and
the existence of some dark matter appears unavoidable.  
A model combining MOND with hot dark matter made of
sterile neutrinos seems to be able to describe
most of the astrophysical phenomenology, from the power spectrum of
the cosmic microwave background anisotropies to 
the dynamics of dwarf galaxies. Whether there exists 
a yet unknown covariant theory that contains 
General Relativity and Newtonian gravity in the weak field limit, 
and MOND as the ultra-weak field limit is still an open question. }

\section{Introduction}
\label{sec:intro}

The dynamical properties of galaxies and larger cosmic structures 
should be described by their mass content if gravity is the dominant
force field. In 1932, Oort noticed a shortage of mass required to
describe the velocity of stars in the solar neighbourhood \cite{oort32}. In 1933, 
Zwicky found the very same problem by applying the
virial theorem to the Coma cluster, based on the radial
velocities of a few galaxies derived from their optical spectra \cite{zwicky33}. 
These early discoveries were revived in the seventies when
the existence of flat rotation curves
in disk galaxies \cite{roberts73} and the requirement of the
dynamical stability of galactic disks \cite{ostriker73,einasto74}
showed that a large fraction of mass, in addition to the observed luminous mass, 
was necessary to describe the dynamics of galaxies.
Ever since, the evidence of missing mass in cosmic
structures has become overwhelming \cite{diaferio08}. 

General Relativity (GR) has encountered such a plethora of successes
in the solar system that the scientific community finds it hard
to suppose that GR, and its Newtonian weak field limit, might fail  
on cosmic scales. It is more natural to suppose that this
missing mass is actually some form of dark matter (DM) that does not emit any
electromagnetic radiation but only acts gravitationally.

This idea meets the demand of particle physics: going beyond the standard model of
particle physics requires the existence of still unknown 
elementary particles, like, for example, supersymmetric particles, axions,
Kaluza-Klein excitations, or sterile neutrinos. These
particles may naturally play the role of the astrophysical 
DM \cite{bertone05}: they formed in the early Universe and their
relic abundance can fill in the fraction of mass that
is required to describe both the internal dynamics of
structures and their formation by
gravitational instability.

In addition, most of these particles have the advantage of being 
non-baryonic, namely, in the astrophysical jargon,
they are neither electrons nor particles made of quarks.
Non-baryonic matter includes neutrinos and the hypothetical 
Weakly Interacting Massive Particles (WIMPs). So far, only
neutrinos have been detected; all other non-baryonic matter
is still hypothetical. 

Being non-baryonic is advantageous
if gravitational instability drives the formation
of the cosmic structure. In fact, in the inflationary
scenario of the standard hot Big Bang cosmology,  small perturbations due to quantum fluctuations
are inflated to cosmic scales by the $\sim 100$ $e$-folding expansion
of the Universe and provide the initial conditions of the matter
density field. However, this scenario is contradicted
by the Cosmic Microwave Background (CMB) anisotropies 
if the matter density field is mostly
baryonic. In fact, the observation 
of temperature anisotropies $\delta T/T= \delta/3$, with $\delta$ the baryonic matter density
fluctuations, in the CMB, which formed by redshift 
$z\sim 10^3$, yields $\delta T/T \sim 10^{-5}$ on $\theta \sim 7^\circ$
angular scales, corresponding to superclusters and larger structures \cite{smoot92}.
 Gravitational instability
yields a growth rate $\propto (1+z)^{-1}$, or slower. Thus,
superclusters would have matter overdensities
$\delta$ of the order of $\sim 10^{-2}$ today, rather than the observed $\delta\sim 1-10$.
Non-baryonic DM decouples from the radiation field much earlier than baryons
and its density perturbations can start growing at the time of equivalence,
when the radiation and matter energy densities are equal.
At the time of the baryon-radiation decoupling, DM perturbations have already
grown to $\delta\sim 10^{-2}-10^{-3}$ on supercluster scales and they will keep
growing to $\delta\sim 1-10$ by today.

Further evidence of the non-baryonic nature of DM is
the abundance of light elements which are synthetized
in the early Universe. Measures of the primordial abundance of deuterium,
for example, which is particularly sensitive to the photon-to-baryon ratio,
implies a baryon density $\Omega_b h^2=0.0214\pm 0.0020$\footnote{We use
the standard parametrization for the Hubble
constant today $H_0=100$~$h$~km~s$^{-1}$~Mpc$^{-1}$.} \cite{kirkman03}, which
agrees with the baryon density  $\Omega_b h^2=0.02229\pm 0.00073$
implied by the CMB anisotropies \cite{spergel07}, yet it
is smaller than the total matter density $\Omega_m h^2= 0.114$ \cite{komatsu11}.
 
By combining all these pieces of evidence, the standard interpretation 
of the observations of the cosmic structure,
from galactic scales to the CMB, suggests that only 17\% of the matter in the Universe
is baryonic, whereas the remaining 83\% is required to be 
non-baryonic cold DM \cite{komatsu11}. The entire amount of matter however yields only
27\% of the matter-energy density required to make the Universe
geometrically flat, as suggested by the CMB power spectrum \cite{debernardis02}.
The missing 73\% can be easily described by a cosmological
constant \cite{bianchi10}, but a large fraction of the cosmological community
is searching for more sophisticated models, some of them including an additional
dark energy (DE) fluid \cite{diaferio08}.

The current standard $\Lambda$CDM model thus contains non-baryonic cold DM and a 
cosmological constant $\Lambda$.   
However, we are in the uneasy situation where only 4.5\% of the matter-energy 
density of the Universe is made of matter we can detect in laboratories on Earth.
The rest is hypothetical (DM) and has unusual properties, like negative pressure 
($\Lambda$ or DE). To avoid this situation, the alternative
solution is to assume that GR breaks down at some scale. A number of extended
or modified theories of gravity have been suggested.
Among them, Milgrom \cite{milgrom83} proposed MOdified Newtonian Dynamics (MOND), an
empirical law 
that modifies Newtonian dynamics and actually goes beyond a simple alternative theory of gravity.
MOND can elegantly deal with
the galaxy scale phenomenology without DM but is less successful
at describing the dynamical properties of galaxy clusters. 
Famaey and McGaugh \cite{famaey11} have recently extensively reviewed this model.
Here, we summarize the successes of MOND and 
mention how the combination of MOND with the existence of sterile neutrinos of
mass in the range 11~eV$-$1~keV can provide a reasonable description of the observed
phenomenology, from galaxy scales to the CMB.

\section{The Acceleration Scale}
\label{sec:acceleration}

In 1983, MOND was originally introduced to explain the observed rotation curves of
spiral galaxies \cite{milgrom83}: the observed velocities of the stars 
are larger than the 
velocities that Newtonian dynamics would predict based on the distribution 
of the luminous matter \cite{roberts73}. As we mentioned above, 
this observational piece of evidence suggested 
the existence of some hidden mass, or DM, responsible
for a stronger gravitational field and hence larger velocities. 
An alternative solution is that, on these cosmic scales, Newtonian gravity
does not hold. Milgrom \cite{milgrom83} went beyond the suggestion
of a modified theory of gravity that becomes relevant beyond
some large length scale. He rather suggested that Newtonian dynamics
breaks down below an acceleration scale $a_0$. 
In the formulation of classical physics, the relation between 
the acceleration $\vec{a}$ 
acting on a point mass and the Newtonian acceleration $\vec{a}_{\rm N}$ is
\begin{equation}
 \mu\left(\vert\vec{a}\vert\over a_0\right) \vec{a} = \vec{a}_{\rm N} \, ,
\label{eq:MOND}
\end{equation}
where $\mu(x)$ is an unknown interpolating function that satisfies the conditions
$\mu(x\to +\infty) = 1$ and $\mu(x\to 0)=x$.\footnote{A year later,
Bekenstein and Milgrom \cite{bekenstein84} showed that, in spherical
symmetry, equation (\ref{eq:MOND}) is equivalent to a modified theory of gravity where
the standard Poisson equation is replaced by
\begin{equation}
\nabla\cdot\left[\mu\left(\vert\nabla\Phi\vert\over a_0\right)\nabla\Phi\right] = 4\pi G\rho\; ,
\label{eq:AQUAL}
\end{equation}
where $\Phi$ is the gravitational potential and $\rho$ the mass distribution.
A more recent modified theory of gravity that reproduces equation (\ref{eq:MOND}) is the
quasi-linear formulation of MOND (QUMOND) \cite{milgrom10}. This theory has the
advantage of involving only linear differential equations and one non-linear algebraic step.
See \cite{famaey11} for a comprehensive review of the various formulations of MOND as
modified dynamics or modified gravity.} To describe 
the observed rotation curves of spiral galaxies, we need to have the acceleration scale 
$a_0\approx 10^{-10}$~m~s$^{-2}$ \cite{milgrom83b}.
Moreover, Milgrom showed that this modification of Newtonian dynamics also drastically
reduces the amount of DM required in groups and clusters of galaxies 
\cite{milgrom83c}.

More importantly, Milgrom's suggestion provided clean predictions \cite{milgrom83b}
that were confirmed in later years. Most notably: (1) the zero point of the Tully-Fisher
relation; (2) the one-to-one correspondence between features in the baryonic distribution 
and the rotation curve; (3) a larger mass discrepancy, when interpreted with Newtonian
gravity, in low surface brightness dwarf spheroidal 
galaxies; and (4) the validity of the 
Tully-Fisher relation for low surface brightness disk galaxies.   

It also became clear very soon that there are different numerical coincidences 
that  might suggest that
$a_0$ is indeed a fundamental quantity (see \cite{bernal11} for a recent
review). For example, it is rather intriguing 
that $a_0$ is related to the Hubble constant $H_0$ and the cosmological
constant $\Lambda$ with the relations $a_0= cH_0/2\pi$ and $a_0^2= c^2\Lambda/2\pi$, 
where $c$ is the speed of light.

\begin{figure}
  \centerline{\includegraphics[scale=0.65]{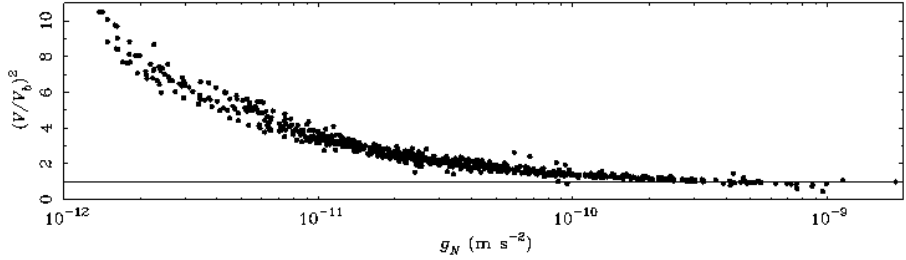}}
  \caption{$(V/V_b)^2$ vs. the Newtonian gravity $g_N = V_b^2/r$, derived
from the baryonic surface density, in 
almost one hundred spiral galaxies. The need for DM appears when 
$(V/V_b)^2>1$. Observations clearly show that this only happens
below a specific acceleration of the order of $10^{-10}$~m~s$^{-2}$. 
Courtesy of B. Famaey and S. McGaugh. Reproduced from \cite{famaey11}, with permission.}
  \label{fig:massdiscrep}
\end{figure}

A compilation of kinematic and photometric data
of disk galaxies very clearly shows the validity of this
{\it ansatz} on the acceleration scale \cite{mcgaugh04}: 
disk galaxies do not require DM 
beyond a given length scale, but rather below a given acceleration
scale. Figure \ref{fig:massdiscrep} shows the ratio between the observed
velocity $V$ and the velocity $V_b$ expected in Newtonian gravity 
from the distribution of the visible matter. The square of this ratio is
proportional to the ratio between the total mass and the visible mass
in a spherical system. Evidently, $(V/V_b)^2>1$ reflects a mass discrepancy
and the requirement of DM; the larger the ratio the larger is
the contribution of DM. Figure \ref{fig:massdiscrep} clearly
shows that the need for DM increases with decreasing Newtonian
acceleration $g_N=V_b^2/r$ which is proportional to the baryonic surface
density. The MOND prediction is thus that low-surface brightness (LSB) 
galaxies are more DM dominated than high-surface brightness (HSB) galaxies
in Newtonian gravity. 
However, the Tully-Fisher relation, that we describe below, 
remains valid for both classes of galaxies.  
In the DM paradigm, it is unclear how the tight correlation 
shown in Figure \ref{fig:massdiscrep} can emerge from the combination
of intrinsically chaotic processes, including the history of the DM halo
formation, the star formation history and feedback processes.

If this acceleration scale does indeed exist, we have two relevant consequences: 
(1) MOND
has to break the Strong Equivalence Principle, which states that all laws
of physics are independent of velocity and location in spacetime, 
and (2) Birkhoff's theorem is not valid. 

MOND is not required to break the Weak Equivalence Principle which asserts the equality
between inertial and gravitational masses. In GR, the Weak Equivalence Principle 
translates into the assumption that all matter fields are coupled 
to the geodesic metric $g_{\mu\nu}$, whereas the Strong Equivalence Principle is guaranteed
by assuming that the same metric $g_{\mu\nu}$ obeys the Einstein-Hilbert action
$S_{\rm EH}=c^4/(16\pi G)\int g^{\mu\nu}R_{\mu\nu}(-g)^{1/2}{\rm d}^4x$, 
where $R_{\mu\nu}$ is the Ricci tensor and $g$ the determinant of the metric.
Therefore, one way to break the Strong Equivalence Principle is to keep the Einstein and 
geodesic metrics distinct.

The non-validity of Birkhoff's theorem clearly complicates
the interpretation of the dynamical properties of self-gravitating systems.
In Newtonian gravity, any object is subject to the gravitational
attraction of all the other objects in the Universe, but in Newtonian linear
dynamics a star in a galaxy or a galaxy in a galaxy cluster are
virtually isolated from the gravitational field felt by the parent
system, unless this latter external field varies across the parent system
and originates the well known effect of tides. In MOND, the external 
field affects the internal motions of the objects even when this field is constant, 
because it is the total acceleration felt by the object that determines
the dynamical, MONDian or Newtonian, regime. We will mention below that
this issue can be relevant in the investigation of the internal
dynamics of elliptical galaxies, that are mostly located in regions
of high galaxy density, and dwarf spheroidal galaxies and globular clusters, that are
satellites of larger galaxies. 

In Sect. \ref{sec:covariant}, we describe an example of how these issues enter the construction of 
a covariant theory of MOND. 

\section{Self-gravitating Systems}
\label{sec:self-grav}

\subsection{Disk Galaxies}

On the scales of galaxies there are a number of observations that were
either predicted or easily explained by MOND. In the DM paradigm, these very same
observations require extreme fine tuning between baryonic and non-baryonic
physics or even yet undiscovered mechanisms.

\begin{figure}
 \includegraphics[scale=0.65]{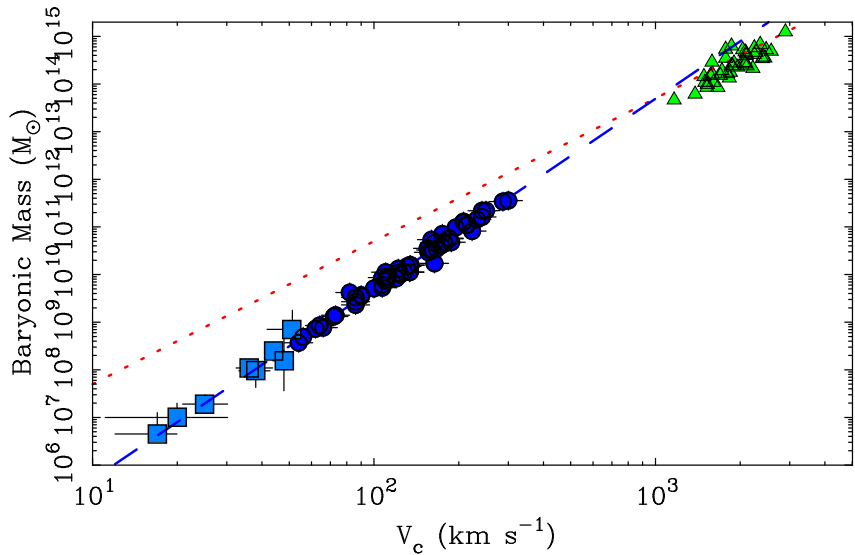}
  \caption{Baryonic mass $M_b$ vs. circular velocity $V_c$ 
in dwarf (squares) and spiral (circles) galaxies \cite{mcgaugh05} and 
  in clusters of galaxies
  (triangles; \cite{reiprich01}).  The blue dashed line is the fit 
to the spiral galaxies alone 
$M_b=50 (V_c/{\rm km\, s}^{-1})^4 M_\odot$. The red dotted line is 
the simplest standard model expectation if all the baryons 
 in each DM halo are identified.  Courtesy of S. McGaugh. Reproduced from \cite{mcgaugh08}, 
with permission.}
  \label{fig:BTF}
\end{figure}

Tully and Fisher \cite{tully77} showed that, in disk galaxies, 
the maximum rotation velocity $v_{\rm rot}$ is proportional
to the galaxy luminosity. MOND predicts that at large radii $a^2/a_0 \sim a_N = GM_b/r^2$ (equation \ref{eq:MOND}) and from $a=v_{\rm rot}^2/r$ we obtain 
\begin{equation}
v_{\rm rot}^4=GM_ba_0 \, ,
\label{eq:TF}
\end{equation}
where $M_b$ is the baryonic mass. Figure \ref{fig:BTF} shows the
measured baryonic mass $M_b$ as a function of the measured circular velocity $V_c$  
for astrophysical systems on different scales, from dwarf galaxies (left bottom
corner) to clusters of galaxies (upper right corner). $V_c$ is 
a characteristic circular velocity measured in the outer region
of the system \cite{mcgaugh08}. The blue dashed line
shows the relation $M_b = 50(V_c/{\rm km\, s^{-1}})^4 M_\odot$.
The MOND prediction clearly agrees with both the observed slope and 
normalization.
In addition, the observed spread is consistent with the uncertainties. This result 
implies that the relation $v_{\rm rot}^4=GM_ba_0$ holds exactly. 

In the standard model, disk galaxies are embedded in DM halos
whose average density within their virial radius is basically
independent of the halo mass. It is thus usual to define $r_{200}$ as
the radius within which the average density is 200 times the 
critical density of the Universe. It follows that for $M_{200}$, the mass within 
$r_{200}$, we have $M_{200}\propto r_{200}^3$ and the circular velocity
$V_c=(GM_{200}/r_{200})^{1/2}\propto M_{200}^{1/3}$. This relation
is shown as the red dotted line in Figure \ref{fig:BTF} . 
The circular velocity is not necessarily identical to the disk rotation
velocity $v_{\rm rot}$, because of the complex interplay between the 
merger history of the DM halo and the star formation history
and energy feedback of the galaxy. 
For example, we could easily recover the correct slope $v_{\rm rot}^4\propto M_b$, 
by assuming that luminosity traces the total mass ($L\propto M_b\propto M_{\rm tot}$) 
and that the density and scale height of the galaxy disk is 
roughly constant in disk galaxies; 
this latter assumption implies $M_{\rm tot}\propto R^2$
where $R$ is the disk size. From $v_{\rm rot}^2=GM_{\rm tot}/R$, we 
correctly derive $v_{\rm rot}^4\propto M_b$. 
A rigorous comparison between
observations and the standard model is not trivial \cite{steinmetz99}, 
but recent analyses, where properly balanced contributions of the various
physical and observational effects are carefully blended, 
seem to bring the $\Lambda$CDM
Tully-Fisher relation in better agreement with observations \cite{desmond12, dutton12}. 

However, in the standard
model, the predicted scatter remains larger than observed, because, unlike MOND, 
we do expect that the galaxy merger and star
formation history mentioned above introduce an intrinsic
scatter. In addition, 
{\it a priori}, we do not have any reason
to expect that (1) the observed relation extends over five 
orders of magnitude in baryonic mass, from dwarf 
galaxies to massive disk galaxies, again with basically no 
scatter, as shown in Figure \ref{fig:BTF}; 
and that (2) the LSB
galaxies, that should presumably have a star formation efficiency lower than 
normal galaxies \cite{wyder09, vanderkruit11}, also perfectly fit into this relation.
This latter result was originally predicted by Milgrom \cite{milgrom83b},
fifteen years before the 
measurements of the rotation curves of LSB galaxies \cite{mcgaugh98}. 

\begin{figure}
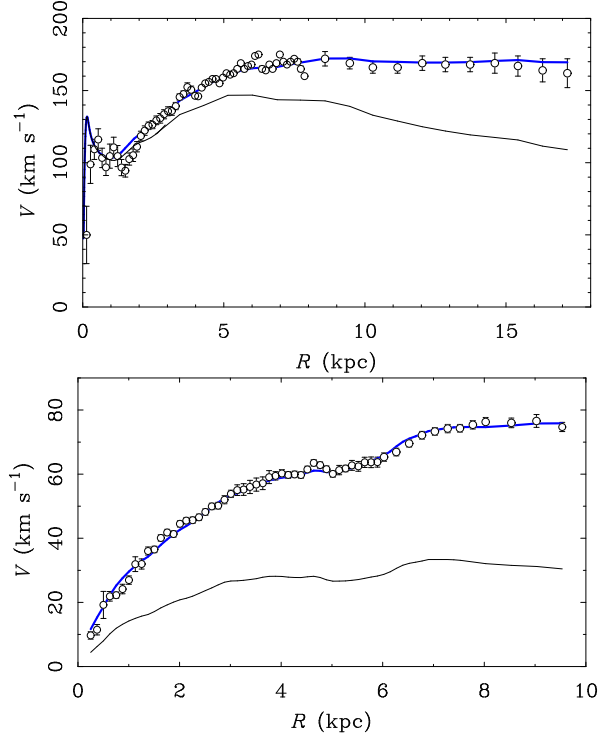

  \centering
   \includegraphics[scale=0.45]{Figure/n6946_LR_mond.ps}\\
   \includegraphics[scale=0.45]{Figure/n1560_LR_mond.ps}
  \caption{Examples of MOND rotation curve fits of a HSB galaxy (NGC~6946, top panel) 
and a LSB galaxy (NGC~1560, bottom panel).
  The black lines show the Newtonian rotation curve expected from the 
observed distribution of stars and gas.
  The blue lines are the MOND fits with best fit stellar mass-to-light ratios in 
the $K$-band $0.37\;M_{\odot}/L_{\odot}$ (NGC~6946) and $0.18\;M_{\odot}/L_{\odot}$ 
(NGC~1560). Courtesy of B. Famaey and S. McGaugh. Reproduced 
from \cite{famaey11}, with permission.}  
  \label{fig:rotcurves}
\end{figure}

MOND goes beyond the description of global properties of disk galaxies. 
Figure \ref{fig:rotcurves} shows the observed 
rotation curves of a HSB 
galaxy and a LSB galaxy (open symbols with error bars). The rotation
curves expected in Newtonian gravity from the distribution of baryonic matter 
(black solid lines) severely underestimate the observations. The
underestimate increases with distance $R$ from the galaxy center and
is larger for the LSB galaxy. In the standard model, this observation
is explained by increasing the DM contribution with increasing $R$ and decreasing galaxy
luminosity. This solution is usually motivated by a lower 
star formation efficiencies at larger radii, as
suggested by extensive surveys of neutral hydrogen in nearby galaxies \cite{leroy08, bigiel10}.

What is more remarkable
is that the small scale features of the rotation curves mirror the
distribution of the baryonic matter in the disk. 
This characteristic is shared by the rotation curves expected in Newtonian
gravity from the distribution of baryonic matter (black solid 
lines in Figure \ref{fig:rotcurves}). 
In the standard model, this property of the total rotation curve 
is unexpected, because this rotation curve should be mostly determined
by the dynamically dominant DM distribution and the baryonic distribution
should play a very minor role. On the contrary, MOND describes the observations
with impressive accuracy (blue solid lines), including the small scale
features of the curves.  

The MOND curves are derived with a single free
parameter, the stellar mass-to-light ratio of the disk, in addition 
to the distance to the galaxy required to determine its length and hence
acceleration scales; the best
fit values of the mass-to-light ratios 
are in perfect agreement with values derived from 
stellar population synthesis models. Moreover, redder galaxies require 
larger mass-to-light ratios, as expected on completely independent 
astrophysical grounds \cite{bell03}. This limited freedom of MOND
should be compared with the standard model that requires two parameters
for the DM halo of each galaxy and a global mass-to-light ratio 
of the galaxy that depends on $R$ and is unrelated to the 
mass-to-light ratios of the stellar populations. 

The agreement shown 
in Figure \ref{fig:rotcurves} is common to 75 nearby galaxies to date \cite{famaey11}.
Among these, an interesting case is NGC~7814: this galaxy is almost perfectly
edge-on and the uncertainties on the rotation curve deriving from the
disk inclination are negligible; in addition, the distance to the
galaxy is accurate to 5\% and is basically not a free parameter any longer. This
galaxy provides a stringent test for MOND: 
high-quality infrared photometric observations \cite{fraternali11} enable, for the
first time, the construction, from an accurate bulge-disk decomposition, 
of a three-dimensional model of the galaxy
whose gravitational potential is inferred by numerically solving the
MONDian Poisson equation. The comparison of the model rotation curve with
the observed one allows the derivation of 
the mass-to-light ratios for both the disk and the bulge components. 
Both ratios are found to be in excellent agreement with the expected values \cite{angus12}.

The data expected in the upcoming GAIA mission and other future surveys 
will provide unprecedented possibilities
to test MOND with the Milky Way dynamics \cite{famaey11}. 
Pioneering work with current data has already
shown that the rotation curve and the surface density of the inner disk of the 
Milky Way are fully consistent with each other within the MOND framework \cite{mcgaugh08b, famaey05}. Similarly, the escape velocity from the solar neighbourhood agrees 
with current estimates if the external gravitational field in which 
the Milky Way is embedded is $a\sim 10^{-2}a_0$: this value is indeed compatible
with the actual field \cite{famaey07}. An additional test involves 
the velocity ellipsoid tilt angle within the meridional
galactic plane. The angles expected in MOND and in the standard Newtonian gravity with DM 
agree with each other and with observations at 
galactic heights  $z=1$~kpc; however, the discrepancy between the predicted angles
in the two models increases with $z$ and the measure of the velocity
ellipsoid tilt angle will thus be a relevant
test to discriminate between the models \cite{bienayme09}.

\subsection{Elliptical Galaxies}

The role played by the acceleration scale is also evident 
in elliptical galaxies, dwarf spheroidal galaxies (dSphs) and globular clusters.

In elliptical galaxies, that are pressure supported systems, Faber and Jackson \cite{faber76} observed a 
relation similar to the Tully-Fisher relation that is valid for disk galaxies: the galaxy
luminosity $L$ correlates with the stellar velocity dispersion $\sigma$
in the galaxy's central region, according to the power law $L\propto \sigma^4$.
If we assume that ellipticals are isothermal spheres that, in 
MOND, have finite mass 
with asymptotically decreasing density $\rho\propto r^\alpha$, with $\alpha=d\ln\rho/d\ln r=-4$, 
we find $\sigma^4=GM_ba_0/\alpha^2$ \cite{milgrom84}. Unlike the Tully-Fisher relation, the observed 
Faber-Jackson relation
is not expected to be exact in MOND, because ellipticals are not strictly isothermal
and their velocity field is not isotropic; the velocity anisotropy
parameter must actually vary with radius to match the observed 
Fundamental Plane of ellipticals \cite{sanders00}. The MOND Fundamental
Plane is actually slightly tilted compared to observations, but this
problem might be removed by including the external field effect \cite{cardone11}. 

Elliptical galaxies pose some challenges to the standard model, because
they should be embedded, like disk galaxies, in DM halos.
X-ray emitting hot gas coronae are expected signatures of a DM halo
and are indeed observed in many early type galaxies \cite{osullivan01},
including isolated ellipticals, like NGC1521 \cite{humphrey12}. 
However, there are cases that are unexpected in the DM paradigm: 
accurate observations, based on planetary nebulae, of the kinematics of the outer parts 
of three ordinary ellipticals show very little evidence, if any, 
of the presence of DM \cite{romanowsky03}, but are in good
agreement with MOND, because the large masses implied by the 
high surface brightnesses indicate that the gravitational field
is in the Newtonian regime $a>a_0$ \cite{milgrom03, tiret07}. 

The kinematics of the outskirts of ellipticals, where $a<a_0$, 
can be probed with spectroscopic observations of their globular cluster systems: 
for example the galaxy NGC4636 in the Virgo cluster is surrounded by 460 globular clusters
with measured velocity and this sample represents one of the largest currently
available. The MONDian predictions agree with the kinematic data and 
NGC4636 also appears to fall onto the baryonic Tully-Fisher relation 
shown in Figure \ref{fig:BTF} \cite{schuberth12}.
 
The X-ray data available for NGC1521 and NGC720 offer a unique test
of MOND in elliptical galaxies: these galaxies are not embedded
in groups or clusters and the X-ray data extend to large radii. 
Similarly to disk galaxies, one
can thus test MOND in the outskirts of galaxies where the 
gravitational acceleration due to the luminous matter is smaller
than $a_0$ and the external field effect is negligible. 
MOND describes the distribution of the baryonic mass in these galaxies
with mass-to-light ratios fully consistent with stellar population
synthesis models \cite{milgrom12}.

\subsection{Dwarf Spheroidals}

At the low mass end of the galaxy mass function, 
dSphs also pose a challenge to the standard model
(see \cite{kosowsky10} for a recent review of dSphs in MOND). 
dSphs have low surface brightnesses and, according to MOND, 
based on what is shown in Figure \ref{fig:massdiscrep} for disk galaxies, 
Milgrom predicted that they should be DM dominated 
and have mass discrepancies larger than ten
when analyzed with Newtonian gravity \cite{milgrom83b}.
This prediction was
impressively confirmed when the first measures of the stellar 
velocity dispersion in the central regions of these galaxies were
available a decade later \cite{milgrom95}. 

More recently, intense observational programs provided velocity
dispersion profiles of the dSphs orbiting the Milky Way \cite{walker07}. This 
piece of information
enables the estimate of the combination of the mass profiles of the galaxies with
the profiles of their velocity anisotropy parameter.  
The most recent detailed dynamical analysis \cite{serra10} 
confirms that the mass-to-light ratios 
in the $V$ band are in the range $1-3$~$M_\odot/L_\odot$, and are
therefore consistent with stellar population synthesis models.
This analysis solved an open issue raised earlier 
\cite{angus08}: unbound stars can contaminate the velocity
dispersion and artificially inflate the estimate of the mass-to-light ratio.
When the unbound stars are properly removed with the caustic
technique \cite{diaferio99, serra11}, Sculptor and Sextans do indeed
show the sensible mass-to-light ratios $1.8$~$M_\odot/L_\odot$ and $2.7$~$M_\odot/L_\odot$
respectively, whereas Carina still
shows a too large $\sim 6$~$M_\odot/L_\odot$. 

This discrepancy might originate from (1) the uncertainties of the luminosity distribution
that is challenging to estimate accurately enough in LSB 
galaxies, and (2) the ellipticity of Carina 
that significantly departs from the spherical symmetry assumed to derive
the mass-to-light ratio. However, this discrepant mass-to-light ratio 
might have a deeper origin due to the specific feature of MOND
that we mentioned in Sect. \ref{sec:acceleration},
the external gravitational field effect:
a star in a dSph, that is a satellite of the Milky Way,  
moves according to both the dSph mass and the
gravitational field exerted by the Milky Way; only if this latter
external acceleration is negligible compared to the acceleration internal to 
the dSph is the dSph mass derived from the stellar velocity dispersion
accurate. Carina is one of the least luminous, and presumably 
least massive, dSphs and one of the closest
to the Milky Way. It is therefore reasonable to suspect that the Milky
Way's gravitational field can play a role in inflating the velocity
dispersion of this dSph. Although this suggestion still awaits
a quantitative confirmation, this same effect appears to be responsible
for the deviation of some dSphs from the expected Tully-Fisher relation
shown in Figure \ref{fig:BTF} \cite{mcgaugh10}.

An additional piece of evidence, which is problematic for the standard model,
is the phase-space distribution of the dSphs that are satellites of the Milky
Way. 
These dSphs are distributed over an extended thin disk whose thickness is between 10 and 
30~kpc and radius $\sim 200$~kpc.
An invoked solution is that these dSphs fell into the MW halo as a small
group of galaxies who kept their orbits correlated \cite{li08}. However, recent
measures of the dSph proper motions indicate that this scenario is
untenable because, according to these measures, 
four dSphs must have fallen in at least 5~Gyr ago \cite{angusdSph11} and 
$N$-body simulations of the standard model show that the
orbit correlation cannot be preserved for such a long time 
\cite{klimentowski10}.

In MOND, dSphs may form as tidal debris during close encounters
of large galaxies \cite{kroupa10} 
and the orbit correlation would thus be a natural consequence of this formation
process \cite{pawlowski11, pawlowski12}. The formation of tidal dwarf galaxies has been
observed in interacting galaxies for the last twenty years \cite{duc12}, 
since the first detection in the Antennae \cite{mirabel92}. The observed stellar velocities
of these systems agree with MOND in the majority of the systems using sensible values of the two 
available free, but constrained, parameters, the mass-to-light ratio and the velocity
anisotropy parameter $\beta$ 
\cite{milgrom07, gentile07}. In the standard model these systems are somewhat challenging 
because, in this case, the observed velocities require a factor of a few more 
mass than the observed luminous mass, despite the fact that
there is no physical reason for these low surface brightness 
tidal dwarfs to drag large amounts of DM \cite{bournaud10}.

The formation of dwarf galaxies as tidal debris is also likely to solve the 
missing satellite problem of the standard model that predicts a factor
of ten more satellites in the halos of large galaxies 
than is actually observed \cite{moore99}.
In MOND, the rate of galaxy encounters is small enough that 
it might provide the right number of satellites with the correct dynamics.
On the contrary, in the standard model, 
a conspiracy of processes regulating the star formation
efficiency is required so that most DM satellites form no stars \cite{guo11}.
In addition, the possible supersonic relative velocity between baryons and 
DM before reionization
might be responsible for the inhibition of the formation of half the expected 
luminous satellites \cite{bovy12}. 
If we take into account the ultra-faint galaxies surrounding the Milky Way, the problem 
can also be partly alleviated, but there would still be a factor
of four too few dwarfs \cite{simon07}. In addition these ultra-faint galaxies 
show indications of tidal disruption, although one would expect
that, with mass-to-light ratios of the order of $1000$~$M_\odot/L_\odot$,
the large DM halo within which they are embedded should be sufficient to screen the stellar
components from the external tidal field of the Milky Way.  
The properties of dwarf galaxies thus remain challenging for the DM paradigm 
\cite{boylan12, hensler12, kroupa12}.

\subsection{Globular Clusters}

The good agreement between the expected dSph dynamics in MOND and  
observations is more intriguing when we consider globular clusters.
These stellar systems and dSphs roughly have the same baryonic mass but 
different surface brightnesses, stellar populations and ages.
In the standard model, globular clusters are devoid of DM, whereas
dSphs are the cosmic structure with the largest fraction of DM, 
with mass-to-light ratios in the range $\sim 10-1000$~$M_\odot/L_\odot$ \cite{walker07, walker12}.
These discrepancies between globular clusters and dSphs are solved
by invoking low formation efficiencies in low mass DM halos \cite{kravtsov05, kravtsov10} 
and two completely different formation processes for the two kinds
of systems. On the contrary, in MOND, the different observed internal velocities
is exactly what we expect from the different surface 
brightnesses (Figure \ref{fig:massdiscrep}) \cite{gentile10}.

Nevertheless there is a system that requires careful consideration.
NGC 2419 is a globular cluster in the outer halo of the
Milky Way with low enough surface brightness
to be in the MOND regime. It is far enough from the gravitational field 
of the Galaxy that the MOND external field effect might not play a relevant role. 
Similarly to the analyses of the dSphs, it is possible to estimate 
the mass-to-light ratio of the cluster from measurements of the projected
velocity dispersion of the stars \cite{ibata11}.
In MOND, the mass-to-light ratio required 
to describe the stellar kinematics 
is a factor of a few lower than expected (the opposite case as for the Carina
dSph). Invoking a variable velocity anisotropy does not
help improve the comparison between the MOND fit and the data, but
non-isothermal polytropic models seem to provide a MONDian description
of the kinematic and photometric observations of this cluster \cite{sanders12a, sanders12b}.
An additional solution is that the life time of the globular cluster
is long enough that mass segregation has already taken place. In this case,
the distribution of RGB and luminous upper main sequence stars used in the Jeans
analysis is expected to be more centrally concentrated than the other
less massive stars of the clusters; therefore the true gravitational potential
of the cluster might be different from the gravitational potential
actually derived with the Jeans equation.
Another possibility is that globulars have non-standard initial
stellar mass functions. More data and more testable globular
clusters are clearly called for. 

Another system that has seemed to be challenging for MOND is 
the five globular clusters orbiting the Fornax dSph. 
Fornax is the most luminous dwarf spheroidal (by about a factor of ten) 
and is the only classical dwarf of the Milky Way with a system of globulars. 
It was suggested that the surprisingly stronger 
dynamical friction due to the dwarf galaxy's low density component of stars in MOND, 
relative to the dynamical friction from the more dense DM halo in 
Newtonian gravity \cite{ciotti04}, 
would cause the globular clusters to lose their orbital angular 
momentum in a period of time much shorter than the Hubble time 
and create a stellar nucleus in Fornax which is not observed \cite{sanchez06, cole12}. 
However, Angus and Diaferio \cite{angus09b} used an orbit integrator 
with accurate mass models of Fornax both in MOND and Newtonian dynamics 
to convincingly demonstrate that the situation is relatively
easy to explain: the globular clusters can orbit for a 
Hubble time as long as their orbits start near the tidal radius. 
This solution does not apply to the most massive globular cluster;
however, this globular cluster is a statistic of one and could have a 
sizeable line of sight distance where the dynamical friction is negligible.

\subsection{Groups and Clusters of Galaxies}
\label{sec:clusters}

The impressive agreement between observations and MOND predictions on 
the scale of galaxies, based on the introduction
of the acceleration scale $a_0$, is not shared by the galaxy cluster data
\cite{the88, gerbal92}.
Clusters do not perfectly obey the relation between 
circular velocity and baryonic mass of equation (\ref{eq:TF}) (green triangles in 
Figure \ref{fig:BTF}) but seem to require more mass than actually observed. 
In fact, in the core of clusters $a>a_0$ and the luminous
matter should be enough to describe the observed dynamics.
The evidence suggests the opposite is true, because 
the amount of observed mass is a factor of two 
too small. The shortage of mass, which is confined to the 
central regions of clusters and progressively disappears in 
the cluster outskirts \cite{angus08b}, is not as large as the factor
of five or larger we have in the standard model, but it clearly 
poses a challenge to MOND. 

\begin{figure}
 \includegraphics[scale=0.65]{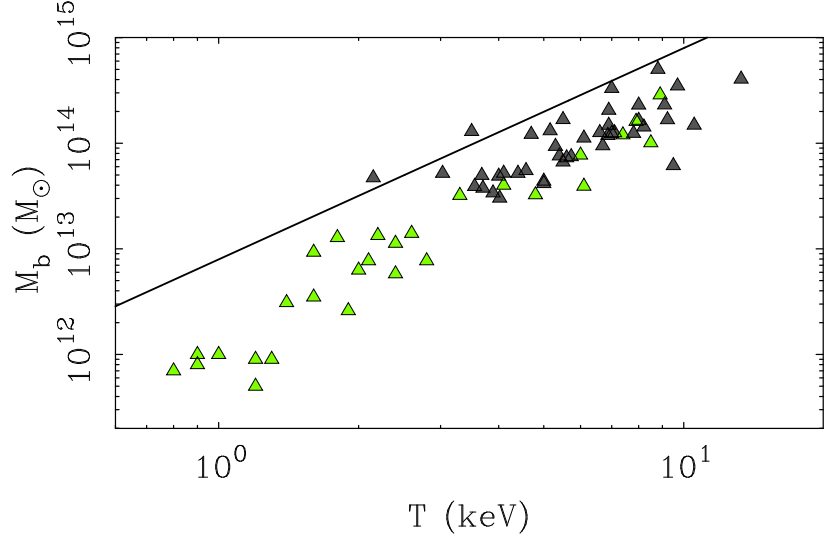}
  \caption{The baryonic mass--X-ray temperature relation for rich clusters (gray triangles \cite{sanders03})
  and groups of galaxies (green triangles \cite{angus08b}).  The solid line is the 
  MOND prediction $M_b \propto T^2$. 
Courtesy of B. Famaey and S. McGaugh. Reproduced from \cite{famaey11}, with permission.}
  \label{fig:cluster-M-T}
\end{figure}

An equivalent, more conventional way to look at Figure \ref{fig:BTF} for clusters
is to consider the relation between the cluster mass and the X-ray temperature 
of the hot intracluster gas. According to the relation 
between mass and circular velocity, MOND
makes a clear prediction for the mass-temperature relation. 
By assuming the same argument used for the Faber-Jackson
relation in elliptical galaxies, the relation 
between the velocity dispersion $\sigma$ of the galaxies
in a cluster and the baryonic mass $M_b$ of the cluster is  
$\sigma^4=GM_ba_0/\alpha^2$, where $\alpha$ is the logarithmic slope
of the mass density profile. The temperature $T$ of the intracluster medium
is a measure of the kinetic energy of the galaxies and $T\propto \sigma^2$.
Therefore, in MOND, we have $M_b\propto T^2$. Figure \ref{fig:cluster-M-T} shows
that the data agree with this scaling relation but not with the normalization, 
because of the mass discrepancy in the cluster cores mentioned above. 

This relation is not at odds with the standard model relation 
$M\propto T^{3/2}$,\footnote{In the standard model,
the cluster virial mass scales as $M_{\rm vir} \propto \rho_c(z) \Delta_c(z) R^3 $, where $R$ is the cluster
size in proper units (not comoving), $\rho_c(z)=3H^2(z)/(8\pi G)$ is the critical
density of the Universe, and $\Delta_c(z)$ is the cluster density in units of $\rho_c(z)$.
A widely used approximation is
\begin{equation}
\Delta_c(z)=18\pi^2 + \left\{ \begin{array}{ll}
 60 w - 32 w^2, &\Omega_m\le 1,\quad \Omega_\Lambda= 0\\
82 w - 39 w^2, &\Omega_m+\Omega_\Lambda=1
 \end{array}\right.
\label{eq:Dcfit}
\end{equation}
where $w=\Omega_m(z)-1$ \cite{bryan98}. Now,
$\rho_c(z)$ scales with redshift $z$ as $\rho_c(z)\propto E^2(z)
= \Omega_m(1+z)^3 + (1-\Omega_m-\Omega_\Lambda)(1+z)^2+ \Omega_\Lambda$.
The cluster size thus scales as $R\propto M_{\rm vir}^{1/3} \Delta_c^{-1/3}(z) E^{-2/3}(z)$,
and the temperature as $T\propto M_{\rm vir}/R \propto M_{\rm vir}^{2/3} \Delta_c^{1/3}(z) E^{2/3}(z)$.} which is known to agree with observations \cite{vikhlinin06},
for two reasons. First, the $M\propto T^{3/2}$ relation is between the X-ray temperature and the total cluster mass, which includes DM, rather than the baryonic mass alone that 
we have in the MOND $M_b\propto T^2$ relation. 
Second, the mass $M$ is 
inferred by assuming valid Newtonian gravity whereas $M_b$ shown in 
Figure \ref{fig:cluster-M-T} is derived with MOND \cite{angus08b}.  
Nevertheless, we can neglect the latter reason and
still see that the two scaling relations are roughly consistent with each other.
In fact, in the standard model, we can write 
$M= M_b/f_b \propto T^{3/2}$, where $f_b$ is the baryon fraction of the
cluster total mass. Therefore, the observed MONDian
$M_b\propto T^2$ should imply $f_b\propto T^{1/2}$. This expectation
is broadly consistent with observations: by combining
X-ray Chandra groups and HIFLUGCS \cite{zhang11} over the temperature
range $[0.6,15]$~keV, Eckmiller et al. \cite{eckmiller11} find $f_b\sim T^{0.79\pm 0.09}$
at $r_{500}$; by limiting the sample to 
clusters with $T$ in the range $[1,15]$~keV, the slope is $0.83\pm 0.42$, but it 
appears to be shallower at smaller radii. 

The apparent mass discrepancy in X-ray bright groups and clusters in MOND \cite{angus08b}, 
clearly shown by the difference 
between the observed and predicted normalizations
of the mass-temperature relation shown in Figure \ref{fig:cluster-M-T}, does
not seem to appear in groups of galaxies 
where the X-ray emission is negligible or absent
 \cite{milgrom98, milgrom02}.
This piece of evidence might suggest that the mass discrepancy 
in MOND can be solved by assuming the presence of undetected baryonic mass
in the form of cold, dense gas clouds that can form and be stable because
of the presence of the ionised hot gas \cite{milgrom08}. 

A different solution is that we might well have
some DM that clusters on cluster scales but not on galaxy scale. 
The piece of evidence that is usually claimed to be difficult to interpret 
without assuming the existence of some form of DM comes
from colliding galaxy clusters. During the collision of two 
clusters, the galaxies and the two halos of DM, 
if DM exists and is collisionless, keep their momentum, 
whereas the intracluster medium is shock heated and slows down.
The DM and galaxy components thus separate from the
gas component. This separation is clearly visibile by 
combining X-ray and optical images of individual 
colliding clusters: for example, the so-called Bullet 
cluster 1E 0657-558 \cite{clowe06}, and the cluster MACS J0024.4-1222 \cite{brada08}.

In the absence of DM, most of the matter resides in 
the gas rather than in the galaxies.
The only way to measure how the total mass is distributed, is to 
derive a map of the gravitational potential with weak gravitational lensing. 
In GR, this procedure is now standard and in the two
systems mentioned above, the mass appears to be concentrated
where the galaxies are and not where the gas is: this is in
striking agreement 
with the presence of some form of collisionless DM dominating
the mass content of the cluster.

Unfortunately, a similar gravitational lensing analysis 
is not trivial in alternative theories
of gravity where the effect of gravity on the light path
is not treated properly. MOND belongs to this group
of theories because it is a classical theory 
and gravitational lensing can only be described by resorting to one 
of the different covariant extensions of MOND. We will describe
the results of this approach in Sect. \ref{sec:covariant}. 
Here, we wish to emphasize that the location of the gravitational peaks
is not an observational fact, but derives
from the assumption of GR's validity. {\it A priori}
it is possible that the non-linearity of a gravity
theory alternative to GR is sufficient to mimic 
the observed gravitational lensing distortion 
with a mass distribution different from that required
by GR. For example, in the covariant MOND theory named TeVeS, 
particular matter distributions
can yield non-zero convergence on sky positions where there
is no projected mass \cite{angus06}. Weyl gravity provides another clear 
non-MONDian example of this phenomenology \cite{pireaux04a, pireaux04b}.

There is a final cautionary comment that cannot be omitted: 
the interpretation of the observed
properties of colliding clusters is far from being clear, because
these systems can also be challenging for the standard model.
The Abell cluster A520 has the same morphology of
the Bullet cluster and MACS J0024.4-1222: two clouds of galaxies on the opposite
sides of a cloud of hot gas \cite{mahdavi07, jee12}. 
However, the GR weak lensing analysis indicates that,
in addition to the two gravitational
potential peaks at the location of the galaxies, 
there is a significant peak where the
gas is. This peak is difficult to explain in the standard model,
because it requires the existence of a massive DM halo devoid
of galaxies, but associated with the cluster hot gas. A number of
possible solutions have been suggested, but none of them appears
to be fully convincing \cite{jee12}.

In addition, colliding clusters appear to have
relative velocities that are unlikely in the standard
model: a system like the Bullet cluster, with a relative velocity
derived from the shockwave of $\sim 4700$~km~s$^{-1}$ \cite{clowe06},
requires an initial infall velocity of the two clusters
of $\sim 3000$~km~s$^{-1}$ \cite{mastropietro08} that has a probability smaller than $3.6\times 10^{-9}$ 
to occur in $\Lambda$CDM \cite{lee10}. On the contrary, the enhanced intensity 
of MOND gravity may naturally produce these large relative velocities \cite{angus08c,
angus11}. Similarly, the 
coherent motion of galaxy clusters on large scales, measured with a technique based on the kinematic
Sunyaev-Zeldovich effect produced by Compton scattering of the CMB photons, 
appears to be challenging for the standard gravitational instability paradigm:
these bulk flows are a factor of five larger than predicted by the standard model
and might require a modification of the theory of gravity, among
other possible solutions \cite{kashlinsky12}.

\section{TeVeS and Gravitational Lensing}
\label{sec:covariant}

MOND, as described in Sect. \ref{sec:acceleration}, is a classical empirical law that 
we would like to recover as the
weak field limit of a covariant theory. This theory should contain GR, that 
excellently describes the gravitational phenomenology of the solar system. 
A covariant theory for MOND is required if we wish to build a cosmological
model and describe the gravitational lensing phenomenology within
the MONDian framework. A covariant theory is thus essential for 
the validation of MOND. 

The attempts to build a covariant theory containing MOND are numerous. A recent overview
of these attempts is provided by Famaey and McGaugh \cite{famaey11}. Here, we briefly
outline one of them, namely TeVeS (see \cite{bekenstein12} for a recent review), whose
Lagrangian contains a time-like vector field and a scalar field in addition
to the tensor field representing the metric, hence the name Te(nsor)Ve(ctor)S(calar).
Halle et al. \cite{halle08} have shown that a very general
Lagrangian for a decaying vector field unifies most of the popular gravity
theories, including quintessence, $f(R)$, Einstein-Aether theories, and TeVeS itself,
among others.  
We stress that TeVeS by no means is the only possible covariant theory that yields MOND in the
weak field limit. Moreover, it is clear that any problem or failure 
belonging to one of these covariant theories are not necessarily
problems or failures of MOND.
 
TeVeS was first proposed by Bekenstein \cite{bekenstein04}. 
At the classical level, MOND introduces an acceleration scale. 
This issue poses an immediate problem for building a generally
covariant theory, because the acceleration scale is played by the affine connection
$\Gamma^\kappa_{\mu\nu}$, that involves the first derivatives of the
metric tensor $g_{\mu\nu}$, and $\Gamma^\kappa_{\mu\nu}$ is not a tensor.
As we mentioned in Sect. \ref{sec:acceleration}, 
one way to bypass this problem is to distinguish between the
Einstein metric $g_{\mu\nu}$, that enters the Einstein-Hilbert action,
and the geodesic metric $\tilde g_{\mu\nu}$, that enters the matter action. 
In GR, the two metrics coincide.
In TeVeS, the two metrics are related by the equation\footnote{In this section, 
we use units where
the speed of light $c=1$.} \cite{sanders97}
\begin{equation}
\tilde g_{\mu\nu}=e^{-2\varphi} g_{\mu\nu} - 2\sinh(2\varphi)U_\mu U_\nu \, ,
\end{equation}
where $\varphi$ is a scalar field and $U_\mu$ is a normalized vector field 
with $g^{\mu\nu}U_\mu U_\nu=-1$. 
Both fields are dynamical. Therefore, the TeVeS action is the sum of three terms: 
the standard Einstein-Hilbert
action, the action term for the scalar field
\begin{equation}
S_\varphi = -{1\over 2k^2 \ell^2 G } \int {\cal F}(k \ell^2 h^{\alpha\beta}\varphi_{,\alpha} 
\varphi_{,\beta}) (-g)^{1/2} {\rm d}^4 x \; ,
\end{equation}
where ${\cal F}$ is an arbitrary positive function, $k$ is a dimensionless coupling constant,
$\ell$ a constant scale length, and $h^{\alpha\beta} = g^{\alpha\beta} - g^{\alpha\mu}g^{\beta\nu}
U_\mu U_\nu$, and the action term for the vector field
\begin{equation}
S_U = - {1\over 32\pi G} \int[Kg^{\alpha\beta}g^{\mu\nu}U_{[\alpha,\mu]}U_{[\beta,\nu]} + 
\bar K(g^{\alpha\beta}U_{\alpha ;\beta})^2 - 2\lambda (g^{\mu\nu}U_\mu U_\nu + 1)] (-g)^{1/2} {\rm d}^4 x\; ,
\end{equation}
where the square brackets indicate antisymmetrisation, $K$ and $\bar K$ are dimensionless
coupling costants and $\lambda$ is a Lagrange multiplier to guarantee the normalization
of $U_\mu$.

TeVeS violates the local Lorentz invariance, because
at each point in spacetime there is a preferred frame in which the time coordinate
aligns with $U_\mu$. The violation of Lorentz invariance 
derives from the invalidity of the Strong Equivalence Principle
anticipated in Sect. \ref{sec:acceleration}. 
Clearly this violation has to be smaller than current experimental bounds.

In the limit $K,\bar K,1/\ell\to 0$, with $k\sim \ell^{-2/3}$, for
quasi-static systems and homogeneous cosmology, TeVeS corresponds to GR.
To recover MOND in the non-relativistic ultra-weak field limit (equation \ref{eq:MOND}), 
we need to choose the function ${\cal F}$.
The arbitrariness of the function ${\cal F}$ makes TeVeS a family of models, rather than 
a single model. 
The function ${\cal F}$ proposed by Bekenstein \cite{bekenstein04} 
yields the MOND acceleration scale 
\begin{equation}
a_0 = {\sqrt{3k}\over 4\pi \ell} \; .
\end{equation}
As we have just mentioned, we recover GR in the limit $K, \bar K \to 0$, whereas to recover
Newtonian gravity in the weak field limit $\ell \to \infty$ suffices.
Therefore, in principle, TeVeS and GR might differ in the strong
field regime \cite{lasky08}.

In the weak field limit, and quasi-static system, the geodesic metric
becomes
\begin{equation}
\tilde g_{\mu\nu}{\rm d}x^\mu{\rm d}x^\nu = -(1+2\Phi){\rm d}t^2 + (1+2\Psi)\delta_{ij}{\rm d}x^i{\rm d}x^j \; .
\label{eq:metric}
\end{equation}
This metric is formally identical to GR, where $\Phi=-\Psi=\phi_N$ and
$\phi_N$ is the Newtonian gravitational potential. 
In this limit, in TeVeS we also have $\Phi=-\Psi$, but $\Phi=\Xi\phi_N+\varphi$, where $\Xi = (1-K/2)^{-1}e^{-2\varphi_c}\sim 1$, with $\varphi_c$ the asymptotic boundary value of $\varphi$. It appears
clear that the scalar field $\varphi$ plays the role that DM plays in the standard model.

Equation (\ref{eq:metric}) enables us to derive the gravitational lensing
equations in TeVeS.
In Sect. \ref{sec:self-grav} we have seen that MOND excellently describes the
dynamics of galaxies but requires some additional DM
on the scales of clusters. One can 
thus anticipate that the description of gravitational lensing 
with TeVeS fits the observations on galaxy scales without
the need of any matter in addition to the observed baryonic
matter, but it does not do as well on cluster scales.

In fact, the TeVeS gravitational lensing
equations applied to the strong lensing regime on galaxy 
scales reproduce the lensing morphology with the observed
baryonic matter alone, both in simple spherically symmetric
models of the lens \cite{chiu06, zhao06} and in models
departing from spherical symmetry \cite{shan08}. These analyses
self-consistently include TeVeS cosmology, because of the cosmological distances of the lenses:
it is easy to get erroneous results
if one uses hybrid theories like MOND combined with GR rather than TeVeS itself \cite{ferreras08}.
In addition, claimed inadequacies of TeVeS \cite{mavromatos09} may be easily healed by assuming
different, but still perfectly reasonable, mass models for the lens \cite{chiu11}.
When analysed properly, to date no lensing system on galaxy scale 
appears to be problematic, including lensed quasars \cite{chen06,chen08,chiu11}.
TeVeS even returns a measure of the Hubble constant consistent with
other independent estimates when its time-delay formula is applied to
lensed variable quasars \cite{tian12}. Nevertheless, the debate on the
adequacy of TeVeS to describe lensing data on galaxy scales is still
open and lively \cite{ferreras12}.

\begin{figure}
\centering
\includegraphics[scale=.65]{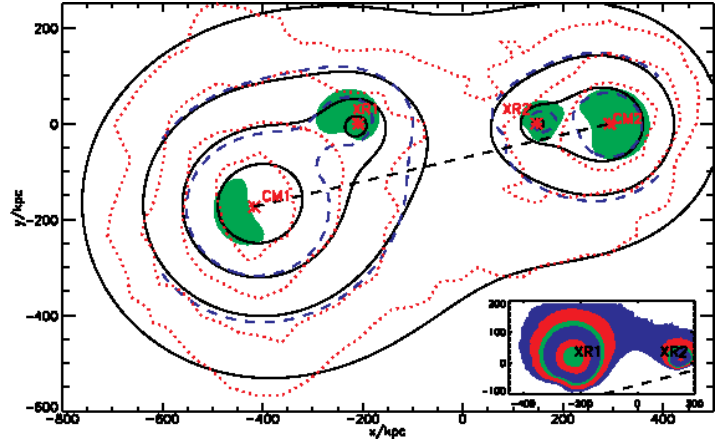}
\caption{The solid black contours show the MOND-ian convergence map of the Bullet cluster \cite{angus07}. The dotted
red contours show the GR convergence map \cite{clowe06}. 
The contour levels are $[0.37, 0.30, 0.23, 0.16]$. The red stars indicate
the  centres of the four potentials used. The blue dashed lines show the contours of surface density [4.8, 7.2]$\times 10^2$~M$_\odot$~pc$^{-2}$ for the MOND standard $\mu$ function.
In the green shaded region the matter density is larger than $1.8\times 10^{-3}$~M$_\odot$~pc$^{-3}$ 
and indicate the clustering of 2~eV neutrinos.
{\it Inset:} The surface density of the gas in the Bullet cluster
predicted by a collisionless matter subtraction method for the standard
$\mu$-function described in \cite{angus07}. The contour levels are $[30, 50, 80, 100, 200, 300]$ M$_\odot$pc$^{-2}$. The origin is [$06^h58^m24^s.38$, $-55^o56^\prime.32$]. Reproduced from \cite{angus07}.}
\label{fig:bullet-lensing}       
\end{figure}

Lensing of clusters returns the mass discrepancy that one encounters 
with the analysis of galaxy kinematics or X-ray emission.
Because of the complexity of the covariant theories of MOND, the lensing
analysis usually assumes spherical symmetry and quasi-stationarity.
Relatively relaxed clusters, where these assumptions are substantially valid, show
the expected mass discrepancies \cite{takahashi07, natarajan08}.
DM halos are required, and it was suggested that DM in MOND could be 
made of neutrinos \cite{sanders03}.
However, the neutrino phase-space density must be smaller
than the Tremaine-Gunn limit we describe below in Sect. \ref{sec:TG}.
The combination of strong and weak gravitational lensing data
shows that the DM central density of MONDian clusters is larger 
than the Tremaine-Gunn limit for neutrino masses smaller than 7~eV \cite{natarajan08}; 
this mass exceeds by more than a factor of three the experimental upper limits on the
neutrino mass and this result suggests that neutrinos are inadequate MONDian DM particles.  
 
The Bullet cluster is usually held as the definitive
proof of the existence of collisionless DM \cite{clowe06}. Unfortunately, 
it is a difficult lens to model in MOND covariant theories because of its
large departure from the assumptions mentioned above of
spherical symmetry and quasi-stationarity. 
Figure \ref{fig:bullet-lensing} shows the results
of one of the first TeVeS analyses of the Bullet cluster \cite{angus07}: the convergence map 
is computed by assuming four spherically symmetric mass distributions
at the location of the two galaxy distributions and the two X-ray
emitting gas clouds. The model agrees with observations if the galaxies
are embedded in additional halos of collisionless matter. Angus et al. \cite{angus07}
concluded that massive neutrinos with 2~eV mass accounting for
two thirds of the total mass of the system is sufficient to reproduce the 
lensing signal. This amount of
additional mass agrees with the results of the dynamical analysis of
other clusters in MOND reviewed in Sect. \ref{sec:clusters}, but, it
became clear later \cite{angus08b, natarajan08}, as we mention earlier,
that the conclusion about the 2~eV mass neutrinos is erroneous. 
The non-linearity of TeVeS does not seem to be conducive to removing the demand for DM
in the Bullet cluster \cite{feix08}. On the other hand, 
the effect of the external gravitational field, that plays 
a role in the internal dynamics of 
dSphs, or the non-trivial features that the additional
TeVeS fields can induce in the lensing phenomena remain to be investigated \cite{ferreira09}.

\section{MOND and Sterile Neutrinos}
\label{sec:MOND+nu}

MOND assumes the existence of an acceleration
scale below which an ultra-weak field limit of the theory of gravity sets in.
From this very simple {\it ansatz}, MOND describes with impressive success the
dynamics of galaxies and smaller systems.
On the scale of galaxy clusters, MOND partly fails (Figure \ref{fig:BTF}) and the observed
kinematics require either a new gravity law or 
the existence of some baryonic or non-baryonic DM.
This failure is bound to be shared by any covariant theory, like TeVeS, 
that is conceived
to yield MOND in the ultra-weak field limit.

For most cosmologists, this failure implies 
the end of MOND as a successful description of the Universe.
However, the DM paradigm currently is far from being satisfactory on galactic
scales \cite{kroupa12}. Therefore, if we are interested in finding a theory that
describes the Universe with the minimum number of assumptions, we
have to consider the possibility that MOND can indeed be a valid description 
of the observed phenomenology and look for possible solutions of its shortcomings
on larger scales. 

Similarly to the standard model, a natural solution 
is the introduction of some form of DM. However, we might have an advantage
over the standard model. In a MONDian model with DM, DM has to be hot enough
to freely stream out of galactic systems, to preserve the
excellent description of the galactic dynamics without DM, but cold enough to cluster
on the scale of galaxy clusters. 
Unlike the hypothetical cold DM particles, we know that an elementary particle
that can play the role of hot DM does exist: the neutrino. 

The properties of neutrinos are currently constrained by various experimental
results. 
In 1979, Tremaine and Gunn \cite{tremaine79}, by considering
the maximum mass density that DM halos made of light leptons
can reach, set a lower limit of $\sim 1$~MeV to the mass of the light leptons
that can make the DM halos of galaxies and clusters of galaxies. 
Below, we briefly review this argument for its relevance to 
the subsequent discussion.

\subsection{The Tremaine-Gunn Limit}
\label{sec:TG}

Neutrinos are collisionless particles, and, according to Liouville's theorem,
the phase-space fluid they form is incompressible. 
In practice, if neutrinos make the DM in self-gravitating systems, like
galaxy clusters, 
this theorem sets an upper limit to the observable coarse-grained phase
space density. 

If self-gravitating systems form by violent relaxation, the neutrino coarse-grained phase-space
density is $f(\vec{x},\vec{v})=f(\epsilon)=f_0 \{1+\exp[\beta(\epsilon -\chi)]\}^{-1}$,
where $\epsilon=\vec{v}^2/2+\phi(\vec{x})$, $\phi(\vec{x})$ is the gravitational potential, 
$\sigma^2(\vec{x})=1/\beta$ is the 1D velocity
dispersion, $\chi$ is the neutrino chemical potential, and $f_0=g_\nu m_\nu^4 h^{-3}$
is the mass phase-space density of an occupied microcell; $g_\nu=2$ is the number of degrees
of freedom, which includes the anti-particles, $m_\nu$ is the neutrino mass, and $h$ the Planck 
constant.

For a non-degenerate neutrino fluid, we have $f(\epsilon)\ll f_0$ which implies
$\beta(\phi-\chi)\gg 0$ and the phase-space density must be smaller than 
the Maxwell-Boltzmann distribution $f_{\rm MB}(\epsilon)=f_0 \exp[-\vec{v}^2/2\sigma^2(\vec{x})]$. 
Therefore, for the neutrino mass density $\rho_\nu(\vec{x})$, we must have 
$\rho_\nu(\vec{x}) \le 4\pi\int_0^{+\infty} v^2 
f_{\rm MB}(\epsilon) dv = f_0 [2\pi \sigma^2(\vec{x})]^{3/2}$. This relation
implies that ${\rm max}\{f\}=f_0\ge
\rho_\nu(\vec{x}) [2\pi \sigma^2(\vec{x})]^{-3/2} $.  
However, clusters form from the relic neutrino background that has the Fermi distribution 
$f(p) = f_0 [1+\exp(pc/k_{\rm B}T)]^{-1}$, with $p$ the momentum and $T=1.95$~K the neutrino 
temperature today. The initial maximum coarse-grained 
phase-space density is therefore ${\rm max}\{f\}=f(p=0)=f_0/2$. 
According to Liouville's theorem, this upper limit cannot increase, and we must thus have 
\begin{equation}
\rho_\nu(\vec{x}) \le {f_0\over 2} [2\pi \sigma^2(\vec{x})]^{3/2}\; .
\end{equation}

For a fully degenerate gas, all the microcells with $v<v_{\rm lim}$ are occupied and
$f(\epsilon)=f_0$, whereas all the microcells with $v>v_{\rm lim}$ are empty 
and $f(\epsilon)=0$. Therefore, in this case, the neutrino mass density is
$\rho_\nu(\vec{x}) \le f_0 4\pi \int_0^{v_{\rm lim}} v^2dv = f_04\pi v_{\rm lim}^3/3 $.
By considering only bound states with $\epsilon={\vec v}^2/2+\phi({\vec x})\le 0$, we have 
${\vec v}^2\le 2\vert\phi\vert\equiv 
v_{\rm lim}^2$, and thus $\rho_\nu(\vec{x}) \le 
f_04\pi \vert 2\phi\vert^{3/2}/3$. By assuming $3\sigma^2=-\phi$,
we obtain $\rho_\nu(\vec{x}) \le f_0 [2\pi \sigma^2(\vec{x})]^{3/2} 4(3/\pi)^{1/2}$,
which implies that the maximum phase-space density is ${\rm max}\{f\}=f_0\ge
\rho_\nu(\vec{x}) [2\pi \sigma^2(\vec{x})]^{-3/2} (3/\pi)^{-1/2}/4$. 
However, applying Liouville's theorem again with the initial 
maximum phase-space density $f_0/2$, 
we obtain  the more stringent upper limit to the neutrino density 
\begin{equation}
\rho_\nu(\vec{x}) \le {f_0\over 2} [2\pi \sigma^2(\vec{x})]^{3/2} 4\left(3\over \pi\right)^{1/2}\; ,
\end{equation}
which is $4(3/\pi)^{1/2}\approx 3.91$ times larger  
than the density upper limit
obtained in the non-degenerate case \cite{kull96,treumann00}. Therefore the non-degenerate case
yields the most restrictive upper limit and is usually called the Tremaine-Gunn limit. 

We can write this limit in astrophysical interesting units as
\begin{equation}
\rho_\nu \le 2.16\times 10^2 \left(m_\nu\over {\rm eV}\right)^4 \left(\sigma \over c\right)^3 {M_\odot \over {\rm pc}^3} 
\end{equation}
or, by considering the relation  $k_{\rm B}T/\mu m_p = \sigma^2$ 
between temperature and velocity dispersion, with
$m_p$ the proton mass and $\mu=0.6$ the mean atomic weight for a fully ionized
gas of solar abundance,
\begin{equation}
\rho_\nu \le 4.64\times 10^{-7} \left(m_\nu\over {\rm eV}\right)^4 \left(k_{\rm B}T\over {\rm keV}\right)^{3/2} {M_\odot \over {\rm pc}^3} \; . 
\end{equation}

\subsection{The Role of an 11~eV Sterile Neutrino}

The mass contribution of the three ordinary (active) neutrinos of the standard
model, $\nu_e$, $\nu_\mu$, and $\nu_\tau$, could, in principle, solve the mass discrepancy in  
MONDian galaxy clusters. At the end of the nineties, laboratory experiments
yielded an upper bound limit to the mass of ordinary neutrinos of 2.2~eV \cite{groom00}.
Therefore, neutrinos with mass 2~eV were proposed as DM in MONDian models \cite{sanders03}.
However, when their contributions to the properties of
astrophysical sources, namely
galaxy clusters, CMB and massive galaxies, are analyzed in details, the ordinary
neutrinos with mass smaller than this limit are shown to be inadequate.

Sanders \cite{sanders03} and Pointecouteau and Silk
\cite{point05} pointed out the relevance of the Tremaine-Gunn limit in galaxy clusters 
in MOND. An extensive analysis of 
26 X-ray emitting groups and clusters of galaxies \cite{angus08b}
considered neutrinos with the maximum mass
allowed by the current upper limits. These neutrinos have a 
Tremaine-Gunn limit that is at least a factor of two 
smaller than the DM density within 100~kpc that is required to describe
the thermal properties of the intra-cluster medium in MOND.
Therefore, either neutrinos are more massive, that is excluded
by laboratory measurements, or they are not the major contribution to the 
MOND DM in clusters.

The three species of ordinary neutrinos with mass in the range $1-2$~eV
also are problematic when we attempt 
to reproduce the CMB power spectrum. These neutrinos 
suppress the third peak by $\sim 25$\% when compared to observations, because of their 
free-streaming imposed by the Tremaine-Gunn limit \cite{mcgaugh04b}.
We can obtain a CMB power spectrum consistent with observations with
these neutrinos in TeVeS if we substantially 
increase the MOND acceleration scale $a_0$ at the time of recombination \cite{skordis06}.
However, this redshift dependence of $a_0$ remains unproved observationally.

Finally, the amplitude of the weak lensing signal around luminous galaxies
($L>10^{11} L_\odot$), extracted from the Red-Sequence Cluster Survey
and the Sloan Digital Sky Survey, suggests mass-to-light ratios larger
than $\sim 10$~$M_\odot/L_\odot$ in the MOND framework \cite{tian09}. 
It thus follows that DM is also required on the scale of massive galaxies.
This DM cannot be made of ordinary neutrinos that are not massive enough
to cluster on these small scales.

The inadequacies of ordinary neutrinos with any mass smaller than the 
current upper limits to properly describe this astrophysical
phenomenology forces us to conclude that ordinary
neutrinos as the MONDian DM are ruled out.  

MOND can still be viable if we resort to hot DM made
of sterile neutrinos. Sterile neutrinos do not have 
standard weak interactions and are right-handed, unlike
the three ordinary neutrinos, and are motivated by a number
of anomalies observed in neutrino experiments (see \cite{abazajian12}
for an extensive review). For example, 
the existence of one sterile neutrino in addition 
to the three ordinary neutrinos  
can elegantly explain the disappearence of 
electron neutrinos from the low energy beam 
measured in short-baseline neutrino oscillation experiments \cite{giunti11a, giunti11,
kuflik12, nelson11}.

If sterile neutrinos are more
massive than ordinary neutrinos, they can have larger
Tremaine-Gunn limits, and thus eventually solve the problems we mentioned above. 
The analysis of the MiniBooNE experiment data does indeed favour
a sterile neutrino mass in the $4-18$~eV range \cite{giunti08}.
This mass range is inconsistent with cosmological constraints in the 
$\Lambda$CDM framework that would require the mass to be smaller 
than 1~eV (see, e.g., \cite{giunti11} and references therein).
However, this mass limit derives from the enhanced
gravity on galactic scales that is required by the standard model 
but not by MOND. Therefore, in MOND, the $4-18$~eV mass range might be
perfectly reasonable.

It is actually quite impressive that a universe with baryonic
matter and one massive sterile neutrino alone reproduces very well the observed
CMB power spectrum if the sterile neutrino mass is $m_{\nu_s}=11$~eV \cite{angus09}, 
a value that is fully consistent with the mass range inferred from the MiniBooNE data. 
In addition, the sterile neutrino mass must be within $\sim 10\%$ of the
11~eV value if we wish to keep the good match with the observed CMB power 
spectrum. This strong constraint is mostly due to the fact that, in this case, the 
contribution of the sterile neutrinos to the density of the Universe
is $\Omega_{\nu_s}h^2 = 0.117$. This value is comparable to the 
contribution of CDM in the standard model that we know to describe
the CMB power spectrum very well. It is important to remind that this analysis
assumes that the sterile neutrinos
are fully thermalised at the time of decoupling; more massive sterile
neutrinos are possible but they must not be fully thermalised \cite{boyanovsky08}. 

\begin{figure}
  \centerline{\includegraphics[scale=0.65]{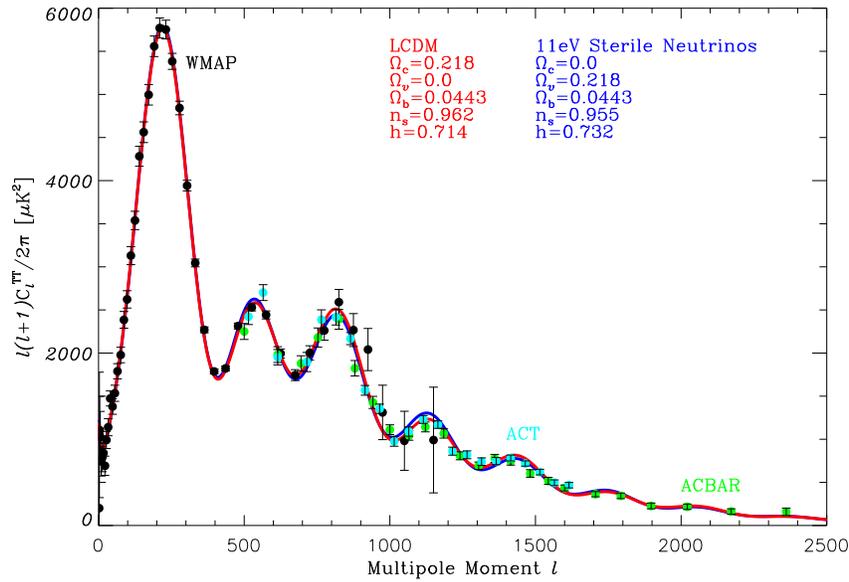}}
  \caption{CMB angular power spectra for a cosmological model 
with baryons and an 11~eV sterile neutrinos (blue line), 
and for the $\Lambda$CDM model (red line). The points show data 
from WMAP year 7 (black), ACT (turquoise) and ACBAR (green). Reproduced from \cite{angus11}.}
  \label{fig:cmb}
\end{figure}

Figure \ref{fig:cmb} shows that the CMB power
spectra of the standard
$\Lambda$CDM model and of a model with baryonic matter and an 11~eV sterile
neutrino are indistinguishable. Note that at the time of recombination,
the average gravitational field is strong enough that the Universe is not  
in the MOND regime, with $a_0$ kept constant to the present day value; 
therefore, the CMB power spectrum can be estimated
with the standard theory of gravity. It is only at later  times 
that the existence of such a massive neutrino
would be problematic for the formation of structure in the standard model.

On the contrary, in MOND this massive sterile neutrino would
be adequate to solve the mass discrepancy on the cluster scale.
In fact, the study of the hydrostatic equilibrium configuration of
30 groups and clusters of galaxies, including the two clumps forming the Bullet cluster 
system, by the analysis of the profiles
of the galaxy velocity dispersion and hot gas 
temperature, shows that the mass density profile of
DM made of 11~eV neutrinos always reaches the Tremaine-Gunn limit in the cluster center \cite{angus10}. 
Some of the dynamical mass must be provided, as expected, by the central galaxies, but the
required amount implies a mass-to-light ratio of $1.2 M_\odot/L_\odot$ in the
K band, in agreement with stellar population synthesis models.
An 11~eV sterile neutrino is also consistent with the straight arc,
originated by a strong lensing effect,
observed in the cluster A2390 \cite{feix10}.

It is remarkable that the 11~eV sterile neutrino required to match 
the CMB power spectrum also solves the completely independent
problem of the mass discrepancy of MONDian clusters
without requiring any additional free parameter or adjustment of the model.
It is also intriguing that the Tremaine-Gunn limit is always reached in the cluster
centers, implying that the dynamical properties of clusters are uniquely set
by the mass of the sterile neutrino. This relation is
completely unkown to the standard DM paradigm, where the
mass of the cold DM particle does not have any role in the dynamical
properties of clusters.  

It remains to be seen whether gravitational instability
in a universe filled of baryonic matter and one species of 
sterile neutrino with 11~eV mass can form the observed
cosmic structure at the correct pace, 
although we remind that the ability to explain the cluster mass discrepancy
does not directly imply that MOND combined with 11~eV sterile neutrinos
can form clusters in a cosmological context. 

The investigation of structure formation in this model requires
an efficient $N$-body code and proper cosmological initial conditions. 
Previous $N$-body simulations of MOND do not consider the presence
of any non-baryonic DM and the adopted initial conditions might not 
be consistent for a universe filled with baryons alone. These previous
$N$-body simulations show an overproduction of cosmic structure
in a baryon-only MONDian universe, in obvious disagreement with
observations \cite{nusser02, llinares08}.  

A first attempt of $N$-body simulation of structure formation 
in a MONDian universe with baryons and an 11~eV sterile
neutrino has been performed with a particle-mesh cosmological $N$-body code 
that solves the modified Poisson equation of the 
quasi-linear formulation of MOND (QUMOND) and with initial conditions appropriate
to DM made of 11~eV neutrinos \cite{angus11}.  
The simulation evolved a box of $512 h^{-1}$~Mpc on a side with $256^3$ particles
from redshift $z=250$ to the present time.

This MONDian hot DM model does indeed produce galaxy clusters with 
the correct order of magnitude of the abundance of observed X-ray clusters, unlike
hot DM models with standard gravity that have structure formation
suppressed on small scales \cite{cen92}. However
the model overproduces the X-ray luminous clusters by a factor of three and underproduces
the low luminousity clusters with $T<4.5$~keV by at least a factor of ten. 
Nevertheless the density profiles
of the simulated clusters are compatible with the observed profiles
of MONDian clusters. In addition, the frequency of relative velocities 
larger than 3000~km~s$^{-1}$ of cluster pairs is large enough to make likely,
unlike the standard model \cite{lee10},
the occurrence of systems like the Bullet cluster. 

Overall the results of this simulation are somewhat unsatisfactory. However,
the simulation has the very poor mass and length resolutions $\sim 10^{12}M_\odot$ 
and $\sim 2$~Mpc, respectively; therefore clustering on small scales is
artificially suppressed and this can partly explain the severe underestimate
of the abundance of low-luminous X-ray clusters. 
The underproduction of low massive clusters could also be quelled
by swapping the 11~eV sterile neutrino for a more massive, up to 1~keV,
sterile neutrinos, because the free-streaming scale decreases with
increasing mass of the DM particle. As we mentioned above, however, these higher
mass sterile neutrinos would not be fully thermalised prior to decoupling. 
The overproduction of massive clusters and supercluster size objects 
actually is a problem that MOND has with any mass of sterile neutrino
or any other DM particle.

In addition, the expansion of the Universe
is assumed to coincide with the standard Friedmann-Robertson-Walker
model with a cosmological constant. It is doubtful that 
a self-consistent relativistic version of MOND will be close enough to 
the standard model to be consistent with current observational limits on the
expansion history of the Universe, but discrepant enough to
yield the correct structure formation rate. Structure formation 
is affected by the acceleration scale
$a_0$ which sets the enhancement of the intensity of the
gravitational field compared to standard gravity. 
Currently, we do not have any observational constraints on 
the redshift dependence of $a_0$, but if $a_0$ is lower
at higher redshift, the overproduction of cosmic structure could
be suppressed and the abundance 
of massive clusters could become consistent with observations. 
However, this feature would suppress galaxy formation and is therefore not ideal.

Clearly, the phenomenological formulation of MOND
is not conceived to yield a self-consistent cosmological model, and 
a covariant model including MOND needs to be implemented to test
its predictions of structure formation robustly. For example, the linear
perturbation theory of the density field in TeVeS has already been
outlined \cite{skordis06a}, but the non-linear evolution of cosmic
structure, with or without a non-baryonic DM component, still needs to be investigated.

\section{Conclusion}
\label{sec:conclusion} 

MOND is based on the {\it ansatz} 
that Newonian dynamics is modified when the gravitational  
field determined by the distribution of baryonic matter  
drops below the acceleration $a_0\sim 10^{-10}$~m~s$^{-2}$ \cite{milgrom83}. 
The observational evidence for the existence
of this acceleration scale has been rapidly accumulating over the last decade.
The success of MOND at describing the dynamics of self-gravitating
systems up to the scale of galaxies with fewer parameters
than the standard DM paradigm is undeniable: 
in the standard model, the additional gravitational force that  
is required by the observed kinematics is supplied by  
the presence of proper amounts of DM, but this solution
requires, unlike MOND, a number of fine tunings and
coincidences to explain the existence of the acceleration
scale that emerges at different length scales. MOND provides 
a far simpler explanation of these observations. 

In addition, on the scale of galaxies, MOND has an impressive predictive power that
is alien to the DM paradigm. 
Numerous observations predicted by MOND were 
confirmed years later. One of the most striking was that, if interpreted
in Newtonian dynamics, the kinematics of LSB galaxies imply that 
DM dominates the dynamics of these systems more than in any
other system in the Universe \cite{milgrom83b}. 

However, on the scale of galaxy clusters, MOND still requires some form of
DM, although not as much as in the DM paradigm \cite{angus08b}. 
Reproducing the power spectrum of the CMB beyond the
second peak also requires some non-baryonic DM \cite{mcgaugh04b}. 

In principle, neutrinos might represent a possible candidate for this non-baryonic
DM. However, current upper limits on the ordinary neutrino mass
show that they would not be massive enough to condense
in the core of clusters and preserve the height of the
third peak of the CMB power spectrum, as required by observations.
   
A viable alternative candidate is a sterile neutrino. 
This particle seems to be required to interpret a number
of anomalous results from neutrino experiments \cite{abazajian12}. A sterile
neutrino with mass in the range 11~eV$-$1~keV would
ensure DM hot enough to stream out of galaxies but cold
enough to cluster on the scale of massive galaxies
and beyond. It is thus in principle capable of explaining
the phenomenology of astrophysical systems preserving
the success of MOND on small scales and, in principle,
the success of the standard model on large scales \cite{angus09}.
Such a model, where we have a modification of gravity
and an exotic DM particle, might be more attractive than
the current standard $\Lambda$CDM model, because of
its elegance on the scale of galaxies, the physical
motivation of the existence of its DM particle from 
experiments on Earth and, eventually, 
the fact that it requires fewer free parameters than $\Lambda$CDM.
However, it remains to be investigated
whether this model can reproduce the full phenomenology of the large-scale structure 
formation and evolution.

Moreover, this model rests on MOND, that 
is still a classical theory, not a covariant theory.
The parent covariant theory that gives MOND in the proper limit
is still unknown. Many possibilities have been proposed, but none 
of them appears yet to be fully convincing.

The exciting conclusion is that, whatever model is going to be
correct, the existence of the acceleration scale, that the data on galaxy
scales indicates with astonishing regularity, has to be explained
naturally by the correct model, unless a number of surprising
coincidences happen to fool us.

\begin{acknowledgement}
We thank Benoit Famaey and Stacy McGaugh for useful suggestions and for
providing us with Figures \ref{fig:massdiscrep}, \ref{fig:BTF}, \ref{fig:rotcurves}, 
and \ref{fig:cluster-M-T}. 
We thank Ana Laura Serra for a careful reading of the manuscript and Luisa Ostorero for
enlightening and encouraging discussions.
AD gratefully acknowledges partial support form INFN grant PD51 and 
PRIN-MIUR-2008 grant \verb"2008NR3EBK_003" ``Matter-antimatter asymmetry,
dark matter and dark energy in the LHC era''. GWA is supported by the Claude Leon
Foundation and a University Research Committee Fellowship from the University of Cape Town.
This research has made use of NASA's Astrophysics Data System.
\end{acknowledgement}
%

%%%%%%%%%%%%%%%%%%%%%%%% referenc.tex %%%%%%%%%%%%%%%%%%%%%%%%%%%%%%
% sample references
% %
% Use this file as a template for your own input.
%
%%%%%%%%%%%%%%%%%%%%%%%% Springer-Verlag %%%%%%%%%%%%%%%%%%%%%%%%%%
%
% BibTeX users please use
% \bibliographystyle{}
% \bibliography{}
%

\end{document}